# Comment on 'Transient response of a wetting film to mechanical and electrical perturbations'

Roumen Tsekov
DWI, RWTH, 52056 Aachen, Germany

A comment on the recent paper of R Manica, J N Connor, L Y Clasohm, S L Carnie, R G Horn, D Y C Chan, Transient response of a wetting film to mechanical and electrical perturbations, Langmuir 2008, 24, 1381.

Recently Manica et al [1] presented theoretical modeling and experimental observations of the previously discovered phenomenon of wimple [2], a rippled deformation of a wetting film during its drainage, and of more complex shapes. These authors claim that they are the first who demonstrate that the wimpling reflects combined action of surface tension, disjoining pressure and hydrodynamics. They make the statements that previous theoretical studies [3, 4] are either incomplete since they "do not consider hydrodynamic effects" or not suitable for the considered experiment since the "treatment of far-field boundary conditions is not appropriate". We comment here on the relevance of the publication of Manica et al [1] to earlier theoretical considerations [3, 4]. Despite critical remarks of Manica et al [1] concerning previous analysis, we demonstrate it does not only fully support their study, but also completely explain their observations and conclusions.

Let us start with the obviously erroneous statement by Manica et al that earlier model [3] does not consider hydrodynamic effects. Comparing Eq. (4) from [3] with Eq. (2) in the paper of Manica et al [1] one obtains an expression for their hydrodynamic pressure in our notations

$$p = 2(p_\sigma - \bar{\Pi})r^2/R^2 + (a - \Delta\rho g)\zeta \qquad (1)$$

where $\bar{\Pi} = \Pi(\bar{h})$. Here the first term is due to the Reynolds hydrodynamic flow between flat surfaces and the second term is a correction for the thickness non-homogeneity $\zeta$ around the average thickness $\bar{h}$. Note that Eq. (1) does not depend explicitly on liquid viscosity. In Stokes hydrodynamics there are two typical problems: finding the pressure distribution generated by a moving object or finding the hydrodynamic flow generated by applied forces. In the latter case, which is relevant to the present comment, the viscosity simply reduces the hydrodynamic velocity but does not affect the pressure distribution. Moreover, any interfacial hydrodynamics contributes only via an effective viscosity of the film [5]. Thus, regardless of the interfacial boundary conditions, the pressure distribution and the thickness inhomogeneity $\zeta$ should have the same spatial distributions in films with different mobility. The latter will affect, however, their evolution via $\bar{h}(t)$. One can see now this similarity by comparing Figs. 3 and 4 from [4], where the surface of a bubble was considered, with Fig. 3 from [1], which imposed unrealistic for a fluid interface stick boundary condition.

By integrating the hydrodynamic pressure from Eq. (1) and the disjoining pressure by taking into account that the average value of $\zeta$ is zero yields the hydrodynamic and disjoining forces

$$F_H = \int_0^R p 2\pi r dr = \pi R^2 (p_\sigma - \bar{\Pi}) \qquad F_D = \int_0^R \Pi 2\pi r dr = \pi R^2 \bar{\Pi} \qquad (2)$$

It follows from Eqs. (2) that the total force $F_H + F_D = \pi R^2 p_\sigma$ depends only on the capillary pressure $p_\sigma$ and it is constant for constant $p_\sigma$ as observed by Manica et al [1]. If the average film thickness obeys the Reynolds law one can calculate the evolution of $\bar{h}$ and substituting it in Eqs. (2) to generate the forces evolution in time. Therefore, the force plots in [1] are just those of the disjoining and capillary pressures. We are now on a position to address the key issue of the paper [1], i.e. the presence of wimples and dimples in the film. The film thickness inhomogeneity $\zeta$ is obviously generated by the flow since films are flat at equilibrium. A scaling estimate of the unknown parameter in Eq. (1) is $a \approx |p_\sigma - \bar{\Pi}|/2\bar{h}$, where the factor 2 accounts for the fact that $\zeta$ is principally spanned between $\pm\bar{h}$. The modulus here reflects the fact that the change of the sign of the driving pressure should reflect in the change of the sign of the deformation $\zeta$ but not in the constant in front. Therefore, using Eqs. (2), our main parameter from [3] can be estimated in the case of strong hydrodynamics as

$$b = (a + \bar{\Pi}' - \Delta\rho g) R^2 / \sigma \approx |F_H| / 2\pi\sigma\bar{h} \qquad (3)$$

Let us now analyze the results of Manica et al by using Eq. (3). One can estimate from their Figs. 2b and 2c that the average thicknesses of the wimple and dimple described are about 100 and 80 nm, respectively. From Fig. 4 in [1] we estimate the corresponding hydrodynamic forces to be about 8 and 3 µN. By using the reported value of the surface tension $\sigma = 0.42$ N/m the corresponding values of the parameter $b$ are estimated from Eq. (3) as 30 and 14, respectively. These values correspond exactly to wimple and dimple according to our qualitative theory [3] showing that dimples and wimples appear at $b < 15$ and $15 < b < 49$, respectively. Let us consider now the dimple presented in Fig. 5 in [1]. From the corresponding hydrodynamic force $F_H \approx -3$ µN and average thickness $\bar{h} \approx 90$ nm it follows that $b \approx 13$, which also agrees with our model. Another example is presented in Fig. 10 showing dimple in 0.1 mM KCl and wimple in 1 mM KCl solutions with average thickness about 30 and 20 nm, respectively. From Fig. 12 one can estimate that the corresponding hydrodynamic forces are $F_H \approx 1$ μN and $F_H \approx 1.5$ μN. Thus, from Eq. (3) it follows that $b \approx 13$ and $b \approx 28$, accordingly. Hence, there is again perfect agreement with our qualitative theory [3]. We should also note that our theory [3] predicts multi-dimpled structures at $b > 49$ being experimentally detected before in different type of experiments [6, 7]. Obviously according to Eq. (3), they are related to extreme hydrodynamic forces, exaggerated by negative disjoining pressure or large film radius, at relatively small average thickness. From this point of view the reported in the study of Manica et al dissipative structures in Figs. 13 and 14 could hardly be seen as a "novel mechanism" as they suggest. Finally, Manica et al [1] mentioned also that the so-called far-field boundary conditions used in [4] are not applicable to their case. However, substituting their film profile beyond the border from Eq. (4) in their Eq. (2) leads to $p = \Pi$, which is nothing more that the classical boundary condition from Eq. (3) in [4]. Moreover, their second boundary condition from Eq. (6), which is not a real one in the physical sense because it just an approximate solution [8] of their Eq. (1), is also valid for our case of a bubble approaching a solid as well as for the case of two bubbles approaching each other [9]. In summary, although the paper by Manica et al [1] contains important high quality experimental data and numerical calculations, it does not report any new physics as they claim. Despite the announcement of the erroneousness of earlier treatment, known models [3, 4] fit these new data well and fully explain their observations.